\documentclass[11pt]{article}
\usepackage{graphicx}
\usepackage{amsmath}
\usepackage{subfigure}
\usepackage{overpic}
\usepackage{hyperref}

\newcommand{\ket}[1]{|#1\rangle}
\newcommand{\bra}[1]{\langle #1|}

\begin{document}
\title{Effects of Entanglement  on  Off-diagonal Geometric Phases}
\author{H. T. Cui, L. C. Wang, X. X. Yi
 \\ \emph{Department of Physics,
Dalian University of Technology, Dalian 116024, China}}
\date{\today}
\maketitle

\begin{abstract}
The effect of entanglement on off-diagonal geometric phases is
investigated in the paper. Two spin-$1/2$ particles in magnetic
fields along the $y$ direction are taken as an example.  Three
parameters (the purity of state $r$, the mixing angle $\theta$ and
the relative phase $\beta$) are chosen to characterize the initial
states. The nodal points at which the usual geometric phases
disappear are calculated and illustrated as a function of the
three parameters. 
\end{abstract}
\maketitle

When a quantum system is transported around a curve $C$ in parameter
space, in addition to the dynamical phase, a geometric phase  can be
developed which is path independent and determined only by the
geometric nature of the Hilbert space. This geometric phase was
first studied by Berry for a quantal system transporting
adiabatically on a closed loop \cite{berry}. The generalization  to
the case of  mixed states was given by Sj\"oqvist \textit{et
al.}\cite{erik} based on Mach-Zender interferometry, and was
experimentally tested in different systems \cite{du}. Besides, the
study on geometric phase has also been extended to non-Abelian case
for pure states\cite{wilczek} and mixed states \cite{singh} as well
as open systems\cite{whitney}. Recently Bhandari has pointed out
that the definition for geometric phase in mixed state fails when
the interference fringes disappear \cite{bhandari}. This  can be
explained as the disappearance (appearance)  of the geometric phase
(off-diagonal geometric phase). The definition of off-diagonal
geometric phase(OP) was first given by Manini \textit{et al.} for
pure states in adiabatic evolutions\cite{manini}, and then was
generalized to non-adiabatic situation\cite{mukunda} and in
mixed states\cite{filipp}. Most recently the off-diagonal geometric
phase was studied in  degenerate case \cite{tong}and in bipartite
systems\cite{yi}.

On the other hand, entanglement as a  property in quantum systems is
proved to be useful in quantum information processing. Furthermore,
it was found that geometric phase may be used to design quantum
logic gates. These facts together give rise to the question of what
the effect of entanglement on the (off-diagonal) geometric phase\cite{ref1}.
We will try to answer this question in the paper.

The off-diagonal geometric phase is complementary to the usual
geometric phase, i.e.,  when the geometric phase is undefined
because the initial state evolves to its orthogonal states, the
off-diagonal geometric phase would  provide the phase information of
state in time evolution\cite{manini}.  In the case of adiabatic
evolution, this can be understood as the appearing of a diabolic point
in the parameter space, at which two orthogonal states have the same
eigenenergy. Thus, the system initially in one of the orthogonal
states  may evolve to another one via this point. In this case the
diagonal phase is undefined and the off-diagonal phase  solely
characterizes the phase change in the evolution and  can be written
as \cite{manini},
\begin{eqnarray}\label{op1}
\gamma_{ij}=\textrm{arg}(\sigma_{ij}\sigma_{ji}),
\end{eqnarray}
where $\sigma_{ij}=\Phi(\bra{\phi_i}U^{\|}(t)\ket{\phi_j})$ with
$U^{\|}(t)$ denoting  a parallel evolution.
$\Phi(z)=z/|z|$ for complex $z \neq 0$.

In what follows, we will discuss the effect of entanglement on the
off-diagonal geometric phase. To this end,   we choose initial state
as
\begin{equation}\label{is}
\rho(0)=\sum _{k=1}^{N} \lambda_k \ket{k}\bra{k},
\end{equation}
where $\ket{k}$ ($k=1,..., M$, $M$ denotes the dimension of the
Hilbert space) are eigenvectors of $\rho(0)$ with corresponding
eigenvalues $\lambda_k$. Because of entanglement, the initial state
Eq. (\ref{is}) cannot be expressed as the direct product of the
subsystem density operators. For two-qubit systems one can use
Peres' condition \cite{peres} to judge whether $\rho$ is entangled
or not. Under parallel evolution $U^{\|}(t)$, one can define GP as
\begin{equation}\label{}
\gamma_{GP}= \text{arg Tr}[U^{\|}(t)\rho(0)]
\end{equation}
In general case, the nodal points, defined as those points in the
parameter space where $\gamma_{GP}$ is undefined, are obtained as
the solutions of
\begin{equation}\label{np}
\text{Tr}[U^{\|}(t)\rho(0)]=0.
\end{equation}
If the equation above is satisfied, we say that $\rho(0)$ and
$\rho(t)=U(t)\rho(0)U^{\dagger}(t)$ are ``orthogonal" to each other,
where $U(t)$ stands for  the time evolution operator for the system.
More ``orthogonal" states  can be found by solving Eq. (\ref{np}).
This  definition of ``orthogonality" between two different density
matrices is universal since it includes not only the case of the
permutation of eigenstates in \cite{filipp}, but also the case of
coherent superposition of eigenstates, that is, for example, the
initial state is in a eigenstate $\ket{k}$, then the ``orthogonal"
state may be $\ket{k'}=\sum_{l} \lambda_l \ket{l}$, in which $l\neq
k $. Consequently, one can define OP as
\begin{equation}\label{op}
\gamma^{(n)}_{OP}=\text{arg}\text{Tr}[U^{\|}(t_{n-1})\sqrt{\rho(0)}\prod_{l=1}^{n-1}U^{\|}(t_{n-1})\sqrt{\rho(t_l)}],
\end{equation}
in which $\rho(t_l)$ is ``orthogonal" to $\rho(0)$ and can be
obtained by solving Eq. (\ref{np}). So far we have presented general
definitions  for nodal points and off-diagonal geometric phases;
exact  expressions for nodal points and off-diagonal geometric
phases depend on the detail of the initial states and the dynamics.

In the remaining  of this paper, we will present a specific
example  to discuss OP and the effect of entanglement. For this
purpose,  we first consider a system consisting  of two non-interacting
spin-$1/2$ particles  in an external magnetic field along the $y$
direction. The Hamiltonian reads ($\hbar=1$)
\begin{equation}\label{h}
H=\frac{\omega_1}{2}\sigma_y^1 + \frac{\omega_2}{2}\sigma_y^2,
\end{equation}
where $\sigma_{y}^i(i=1,2)$ are  Pauli operators for the particle
$i$, $\omega_{i}(i=1,2)$ represent  the procession frequency. We
notice that the off-diagonal item in the Hamiltonian Eq. (\ref{h})
can lead to the transition between different eigenstates of
$\sigma_z^i$ (i=1,2). So under the time evolution governed by $H$ in
Eq. (\ref{h}), a state may evolve to its orthogonal states and hence
the corresponding GP is undefined. Suppose that the initial state
takes the form,
\begin{equation}\label{rho}
\rho(0)=\frac{1-r}{4}I_{4}+r\ket{\Phi}\bra{\Phi},
\end{equation}
where $r\in(0,1]$ determines the purity of the mixed state. And we
take  $\ket{\Phi}$   from the following states,
\begin{eqnarray}\label{rho1}
\ket{\varphi}&=\cos\theta e^{i\beta}\ket{11}_{1,2}
                      +\sin\theta e^{-i\beta}\ket{00}_{1,2} \nonumber\\
\ket{\psi}&=\cos\theta e^{i\beta}\ket{10}_{1,2}
                   +\sin\theta e^{-i\beta}\ket{01}_{1,2}
\end{eqnarray}
where the mixing angle $\theta,$ and the relative phase $\beta$
together determine the degree of entanglement.
$\ket{1(0)}_{i}(i=1,2)$ is the eigenstate of the Pauli operator
$\sigma^i_{z}$ (i=1,2). When $\theta=\pm\pi/4$ and $\beta=0$, the
states Eq. \eqref{rho1} are the well known Bell states and
correspondingly Eq.(\ref{rho}) gives the Werner state \cite{werner}.
The Werner state plays an important role in quantum information
processing, in particular in quantum communication via noisy
channels \cite{bennett} and in the quantum distillation
scheme\cite{bennett1}. This initial state Eq. (\ref{rho}) includes
all possible cases, such as pure or mixed states and maximal or
non-maximal entangled states. One should note that these initial
mixed states are triplet-degenerate.

Now we are in a position to discuss OP. Recently an experimental
scheme for verifying second order OP($n=2$ in Eq. (\ref{op})) has
been proposed\cite{filipp} based on the Franson-type interferometer,
in which an entangled photon  pair is produced by using
spontaneous-down-conversion\cite{kwait}. This scheme can be used to
prepare the initial state Eq. \eqref{rho}. First let us study the
nodal structure for GP given by Eq. (\ref{np}) under parallel
transportation. The parallel transportation can be realized by
imposing\cite{singh}
\begin{equation}\label{pt}
U^{\|}(t)= U(t)V(t),
\end{equation}
where $U(t)$ is the unitary time-evolution operator and the
elements of the blocked matrix  $V(t)$ are defined in this model as
\begin{eqnarray}
V_{\mu\nu}&=&\bra{\mu}e^{it \sum_{\mu', \nu'}\bra{\mu '}H\ket{\nu '}\ket{\mu '}\bra{\nu '}}\ket{\nu},
\ket{\mu}, \ket{\nu} , \ket{\mu '}, \ket{\nu '}\in\{degenerate\ \  subspace\} \nonumber\\
V_{kk} &=& e^{i \bra{k}H\ket{k}t}, \ket{k}\in\{the\  remaining \ space\},
\end{eqnarray}
where $\ket{\mu}, \ket{\nu}, \ket{\mu '}, \ket{\nu '}, \ket{k}$  are the eigenstates of $\rho(0)$,
and the interference terms between the degenerate space and the other space are set to be zero
 in order to keep the parallel transport in the degenerate space.
We would like to note that the parallel transport operator
$U^{\|}(t)$ depends on the initial  state. So, different initial
state leads to different nodal structure. The dependence of the
nodal structure on the parameters of $r$, $\theta$ and $\beta$  can
be obtained by solving Eq. (\ref{np}). In order to simplify the
tedious expression, we set $\omega_1=\omega_2=\omega$ and $\omega t=
\pi$. The results then are reduced to
\begin{eqnarray}\label{gnp}
\text{Tr}[U_{1}^{\|}(t)\rho_{1}]=\frac{1+3r}{4}\cos2\beta\sin2\theta
 + \frac{1-r}{4}[1 - (1 + \cos2\beta\sin 2\theta)\cos\frac{\pi}{2}\sqrt{2 + 2\cos2\beta\sin2\theta}], \nonumber\\
\text{Tr}[U_{2}^{\|}(t)\rho_{2}]=-\frac{1+3r}{4}\cos2\beta\sin2\theta
 +  \frac{1-r}{4}[1 - (1 - \cos2\beta\sin 2\theta)\cos\frac{\pi}{2}\sqrt{2 - 2\cos2\beta\sin2\theta}],
\end{eqnarray}
where
\begin{eqnarray}
\rho_{1}&=&\frac{1-r}{4}I_{4}+r\ket{\varphi}\bra{\varphi}, \nonumber \\
\rho_{2}&=&\frac{1-r}{4}I_{4}+r\ket{\psi}\bra{\psi},
\end{eqnarray}

\begin{figure}[t]
\begin{overpic}{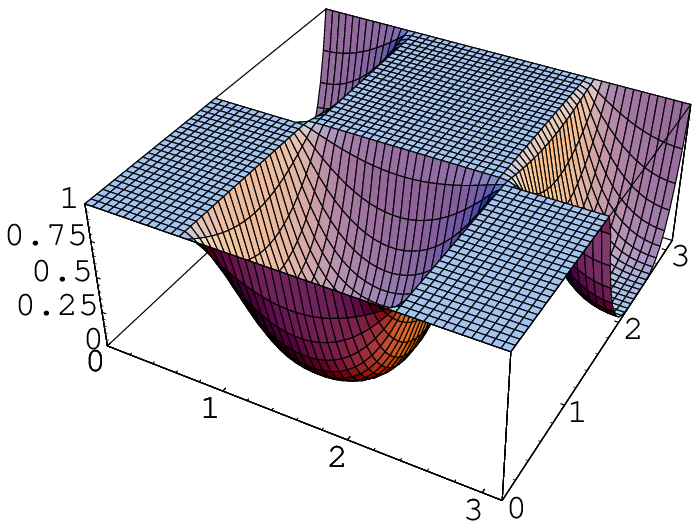}
\put(10, 75){\textbf{(a)}}
\put(35,10){$\beta$}
\put(90, 20){$\theta$}
\put(5, 50){$r$}
\end{overpic}s
\begin{overpic}{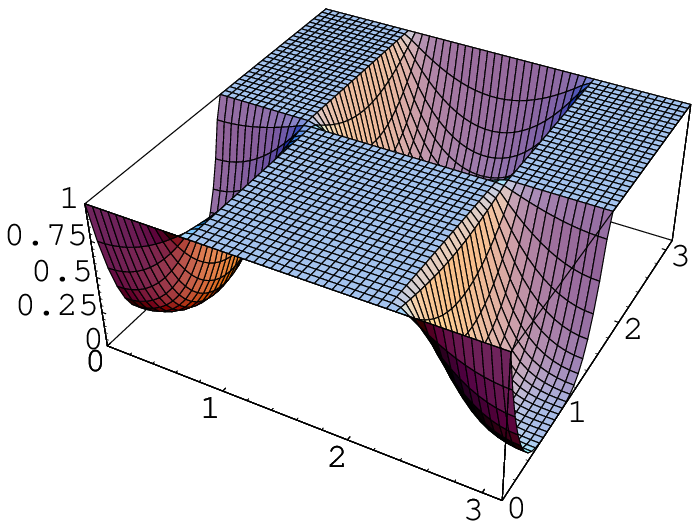}
\put(10, 75){\textbf{(b)}}
\put(35,10){$\beta$}
\put(90, 20){$\theta$}
\put(5, 50){$r$}
\end{overpic}
\caption{\label{g1}The nodal structure of GP for mixed states
$\rho_1$ (a) and $\rho_2$ (b) vs $\beta$[Arc], $\theta$[Arc] with
$\omega_1t=\omega_2t=\pi$.}
\end{figure}

\begin{figure}[t]
\centering
\begin{overpic}{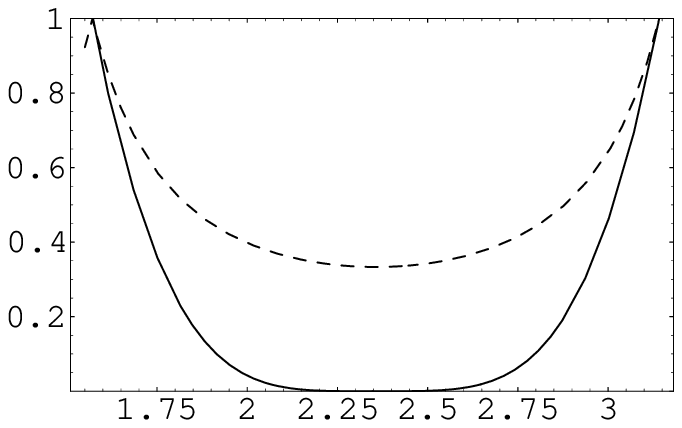}
\put(-5, 35){$r$}
\put(50,-5){$\theta$}
\end{overpic}
\caption{\label{g11}Comparison between the nodal structure and
entanglement in the initial state. The dashed line denotes
$r=\frac{1}{1+2|\sin2\theta|}$. We only draw for $\rho_1$ with
$\beta =0$ and the other parameters are the same as those in
Fig.\ref{g1}. Similar conclusion can also  be found for $\rho_2$. }
\end{figure}

The results show that there exist some common nodal points for the
two initial states with $r=1$ and $\cos2\beta=0$ or $r=1$ and
$\sin2\theta=0$, which indicates that the common nodal points lie in
separable pure states. Note that  the nodal points given by Eqs.
(\ref{gnp}) are a function of $\theta, \beta$ with period of  $\pi$,
then we may plot the nodal points only within $[0,\pi]$ for both
$\theta$ and $\beta$. The detailed dependence of the nodal structure
on the three parameters was shown in Fig.\ref{g1}. It is clear that
there are more nodal points in mixed states than that in pure
states. Furthermore  we find that the nodal  points appear only with
$0< r \leq \frac{1}{1+2|\sin2\theta|}$ (see Fig.\ref{g11}), that is
exactly the condition for $\rho_{1,2}$ being separable   states.
However, we find that when the initial state is a Werner state, no
nodal points exist in the parameter space for any $r\in(0, 1]$.

Next let us study the two-indexes OP ($n=2$). According to the
definition of Eq. (\ref{op}), we may write the two-indexes
off-diagonal geometric phase  as
\begin{equation}\label{}
\gamma^{(2)}_{OP}=\text{arg}\text{Tr}[U^{\|}(t)\sqrt{\rho(0)}U^{\| '}(t)\sqrt{\rho(t)}],
\end{equation}
where $\rho(t)=U(t)\rho(0)U^{\dagger}(t)$, and  parallel evolution
$U^{\|'}(t)$ is  different from the previous one. The same
conditions, namely $\omega_1=\omega_2=\omega$,  were chosen to
simplify the expression.  With this assumption, the two-indexes OP
($n=2$) is given by
\begin{eqnarray}\label{onp}
\gamma^{(2)}_{OP}(\rho_{1,2})=\textrm{arg}[\frac{1+r}{2}+\frac{1-r}{2}\cos\pi\sqrt{2\pm\cos2\beta\sin2\theta} ]
\end{eqnarray}
Obviously the value of $\gamma^{(2)}_{OP}(\rho_{1,2})$ is always
zero when the diagonal geometric phase is undefined, independent of
the initial state. This property is very different from the diagonal
geometric phase, which depend not only on the initial state, but
also on the dynamics of the system.  It furthermore reflects the
topology of the off-diagonal geometric phase, as the value of OP
 depends on the degeneracy \cite{manini}. In fact since the
nodal structure for diagonal geometric phase appears only for
unentangled state and there is no intra-coupling in our model, then
OP  is  the sum over that of the two particles.

\begin{figure}[t]
\begin{overpic}{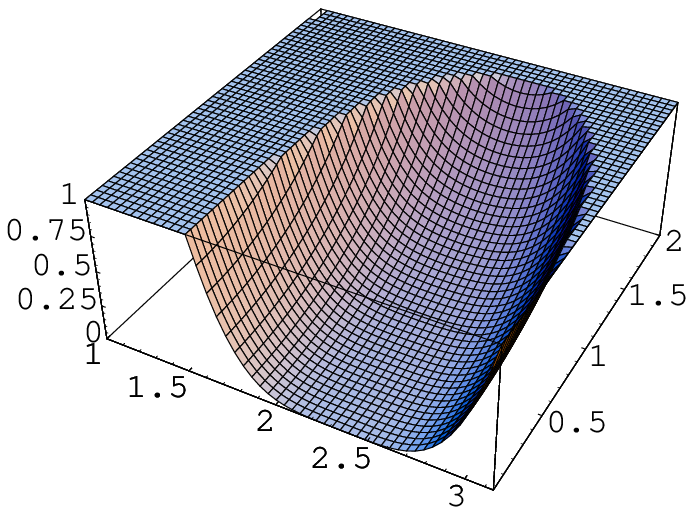}
\put(10, 75){\textbf{(a)}}
\put(35,10){$\theta$}
\put(90, 20){$J$}
\put(5, 50){$r$}
\end{overpic}
\begin{overpic}{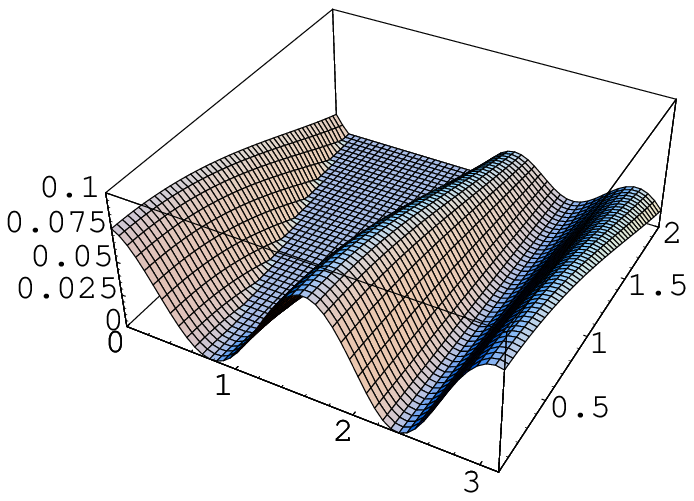}
\put(10, 75){\textbf{(b)}}
\put(35,10){$\theta$}
\put(90, 20){$J$}
\put(5, 50){$r$}
\end{overpic}
\caption{\label{g2}The nodal structure of GP for mixed states
$\rho_1$ with  inter-subsystem couplings vs $\theta$[Arc]and the
rescaled coupling constant $J= g/\omega $. (a) corresponds to the
real part of Eq. \eqref{np} and (b) corresponds to the imaginary
part of Eq. \eqref{np}.  $\omega_1t=\omega_2t=\pi$ and $\beta=0$
were chosen for this plot. }
\end{figure}

So far we only discuss the nodal structure of geometric phase
without inter-subsystem couplings. Another question is how the nodal
structure is affected by the inter-subsystem couplings. For this
purpose we consider the Ising-type interaction,
\begin{equation}\label{}
H_I=\frac{g}{4}\sigma_z^1\sigma_z^2.
\end{equation}
In general the interaction tends to destroy the degeneracy or the
level (avoided) crossing  and we guess that the nodal structure
should be compressed because of the couplings. After calculations we
found that this is the casae, i.e., the intra-coupling tends to
destroy the nodal structure and the nodal points only appear for
special initial states with certain coupling constant. We do not try
to list the results of our calculation because of the complicated
expressions, the selected numerical results were shown in Fig.
\ref{g2}.
From the figure we see that because of the inter-subsystem
couplings, Eq. \eqref{np} is a complex number and hence in order to
determine the nodal points, we have to let the real and imaginary
part of the complex number to be zero simultaneously. Different from
the free case, the nodal points appear only for the special values
of $\theta$, displayed in Fig. 4 and only exist for weak couplings
($J\leq 0.28$). Moreover we find that the separable condition for
the initial state is required too. Besides  when the initial state
is a Werner state, there is no nodal point appearing. The two-index OP
has been calculated and our calculations show that the values of OP
are not always  zero or $\pi$, this result  comes from the
intra-coupling and also depends on the degeneracy. Similar
conclusions can be found for $\rho_2$.
\begin{figure}[t]
\centering
\begin{overpic}{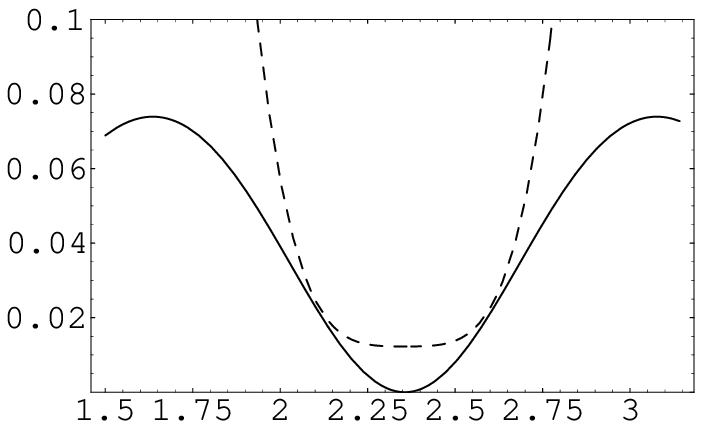}
\put(-5, 35){$r$}
\put(50,-5){$\theta$}
\end{overpic}
\caption{\label{g22}Detailed demonstration for the Fig.\ref{g2}.
We have chosen $J=0.28$ and the other parameters is the same as that
in Fig.\ref{g2} . The dashed line corresponds to (a)  and solid line
corresponds to (b) in Fig.\ref{g2}. With the decrease of $J$, the
dashed line moves downward.}
\end{figure}

In conclusion, we have discussed the entanglement effect on the
off-diagonal  geometric phase in mixed states by a general example
and some novel results are presented. First, because of entanglement
and mixture of the initial states, the nodal structure for GP have
been changed greatly, as displayed in Fig. \ref{g1}. We found that
the nodal points only appear in region $0<r\leq
\frac{1}{1+2|\sin2\theta|}$ as displayed in Fig.\ref{g11}, which
means the initial states are separable. Furthermore when the initial
state is a Werner state, we found there is no nodal point for any
$r\in(0,1]$. The two-indexes OP for the nodal points  are shown to
be  zero,  independent of the initial states. We also extend this
discussion to the case with  inter-subsystem couplings. The results
show that because of the coupling the nodal points for GP have been
compressed greatly and only appear for some special initial state
and small coupling constants. Besides we find that similar to the
free case, there is no nodal point for the Werner state, implying that
the Werner state is robust against the perturbation.

This work was supported by NSF of China under grants 10305002 and
60578014.


\begin{thebibliography}{99}
\bibitem{ref1} The effect of entanglement on the geometric phase
was studied already \cite{erik2,tong2}, however the effect on
the off-diagnoal geometric phase in particular for mixed states
remains untouched.
\end{thebibliography}
\end{document}